\documentclass[10pt,pra,aps,twocolumn,showpacs,superscriptaddress]{revtex4}
\usepackage{graphicx}
\usepackage{amsmath,amssymb,revsymb}
\usepackage[colorlinks=true,citecolor=blue]{hyperref}

\begin{document}

\title{Cold three-body collisions in hydrogen-hydrogen-alkali atomic system}
\author{Yujun Wang}
\altaffiliation{Present address: JILA, University of Colorado, 440 UCB, Boulder, Colorado, 80390, USA}
\affiliation{Department of Physics, Kansas State University, Manhattan, Kansas, 66506, USA}
\author{J.P. D'Incao}
\affiliation{JILA, University of Colorado and NIST, Boulder, Colorado, 80309-0440, USA}
\author{B.D. Esry}
\affiliation{Department of Physics, Kansas State University, Manhattan, Kansas, 66506, USA}

\begin{abstract}
We have studied hydrogen-hydrogen-alkali three-body systems in the adiabatic hyperspherical representation. 
For the spin-stretched case, there exists a single $X$H molecular state when $X$ is one of the bosonic alkali atoms: $^7$Li, $^{23}$Na, $^{39}$K, $^{87}$Rb and $^{133}$Cs. 
As a result, the {\em only} recombination process is the one that leads to formation of $X$H molecules, H+H+$X$$\rightarrow$$X$H+H, and such molecules will  be stable against vibrational relaxation. 
We have calculated the collision rates for recombination and collision induced dissociation as well as the elastic cross-sections for H+$X$H collisions up to a temperature of 0.5 K, 
including the partial wave contributions from $J^\Pi$=$0^+$ to $5^-$. 
We have also found that there is just one three-body bound state for such systems for $J^\Pi$=$0^+$ and no bound states for
higher angular momenta. 
\end{abstract}
\maketitle

\section{introduction}
In the last decade, studies of three-body collisional processes have attracted tremendous attention due to their great relevance for the rapidly growing field of cold and ultracold atomic gases~\cite{FeshbachChin}.
In such systems, three-body recombination and elastic and inelastic atom-molecule collisions are of particular interest.

Three-body recombination is a scattering process where three free particles collide, with two of them binding to form a molecular state, 
converting the binding energy into the relative kinetic energy of the atom and molecule produced.
Three-body recombination, therefore, is important generically as it can shed light on binding in nuclear and chemical reactions.
In ultracold atomic gas experiments, three-body recombination can lead to huge losses near a Feshbach resonance~\cite{Ketterle,Jin,Grimm} 
and has been studied extensively to understand the lifetime and the stability of the gas samples~\cite{Fedichev,MacekRecomb,EsryRecomb,Badaque,Weber,Roati}.

Elastic atom-molecule collisions are crucial for determining the dynamics of ultracold atom-molecule mixtures at the mean-field level, and inelastic atom-molecule collisions
have a big impact on the lifetime of Feshbach molecules in such systems~\cite{DimerJochim,DimerColChin,DimerRegal2004}.
Furthermore, in the regime of large two-body $s$-wave scattering length $a$, achieved near a Feshbach resonance, three-body collisional processes
show universal scaling behavior with $a$ as result of Efimov physics~\cite{Efimov,MacekRecomb,EsryRecomb,Badaque,Braaten,BraatenReview}.
These universal aspects have been observed experimentally in recent years~\cite{GrimmRecomb,AtomDimer,ThreeFermions1,ThreeFermions2,KRb,EfimovSpec,ThreeLi7} 
and confirm our understanding of three-body universal properties~\cite{BraatenReview}.

Ultracold three-body collisions, however, are only universal when the dynamics are predominantly determined by the long-range behavior of the atom-atom 
scattering wave function. When a system behaves universally, the complicated atom-atom interaction can thus be replaced with a much simpler model designed to reproduce the long-range wave function. 
It is in this context that the adiabatic hyperspherical representation has been applied to the calculation of three-body recombination. 
While it has proven very useful for getting deeper insights into this process, calculating recombination for chemically important species using realistic interactions requires substantial further technical 
development, but does not pose fundamental difficulties. 

Nevertheless, there are a few realistic systems that are sufficiently simple that recombination calculations are possible with the tools already available. For instance, recombination of helium atoms, for which there is a single $^4$He$_2$
ro-vibrational bound state, has been studied within the adiabatic hyperspherical representation~\cite{SunoHe1,SunoHe2}. Parker, {\it et al.} have studied the 
Ne+Ne+H system, calculating the $J^{\Pi}$=$0^+$ partial wave contribution to recombination and collision induced dissociation rates~\cite{Parker}. 
The studies done by Suno {\it et al.} for three helium atoms~\cite{SunoHe1,SunoHe2} 
and for He+He+Alkali systems~\cite{SunoHeAlk} have included higher partial wave contributions in order to calculate 
 the total recombination rate up to temperatures of at least 10~mK. These results
are relevant to the buffer gas cooling technique used in cold and ultracold experiments~\cite{BufferGas,BufferCool}, where three-body recombination can lead to
dramatic losses.

In this paper, we study three-body processes involving  two hydrogen atoms and one alkali atom. 
If all atoms are spin-stretched, these systems are amenable to calculation since 
 the H+H interaction has no bound state and $X$+H has only a single $s$-wave bound state for all the alkali species, $X$, considered here~\cite{Kolos,Bromley,Cote,Cote1}: $^7$Li, $^{23}$Na, $^{39}$K, $^{87}$Rb and $^{133}$Cs.
In thermal Alkali-hydrogen mixtures,
the recombination rate $K_3$ for the process H+H+$X$$\rightarrow$$X$H+H is 
related to the density of the alkali atoms by
\begin{eqnarray}
\frac{d n_{X}}{d t}=-K_3 n_{\mathrm{H}}^2 n_X
\end{eqnarray}
where $n_H$ and $n_X$ are the densities of the H atoms and alkali atoms, respectively. Note that there is a $2!$ reduction in the 
rate if the hydrogen atoms are in a condensate~\cite{Kagan1985}.
The three-body calculations are thus simplified by having only one recombination channel, but the presence of sharp avoided crossings in the three-body potentials
makes the calculations a challenge in the adiabatic representation. 
These sharp crossings are between different
families of adiabatic potentials corresponding to H+H and $X$+H. 

We will use atomic units throughout unless specified otherwise.  It is also convenient to convert
energies to temperature units by dividing by Boltzmann's constant $k_B$, i.e. 3.17$\times$10$^{-6}$~a.u.=1~K.

\section{method}
\label{method}
After separating the center-of-mass motion, the relative motion of the three particles can be represented by the mass-scaled Jacobi vectors 
$\boldsymbol{\rho}_{12}$ and $\boldsymbol{\rho}_{1,23}$~\cite{Delves}:
\begin{eqnarray}
\boldsymbol{\rho}_{12}&=&(\boldsymbol{r}_1-\boldsymbol{r}_2)/d,\\
\boldsymbol{\rho}_{12,3}&=&d(\boldsymbol{r}_3-\frac{\boldsymbol{r}_1+\boldsymbol{r}_2}{2}),
\end{eqnarray}
where $\boldsymbol{r}_1$, $\boldsymbol{r}_2$, and $\boldsymbol{r}_3$ are the lab-frame position vectors of the two hydrogen atoms with mass $m_H$ and the alkali atom
with mass $m_X$, respectively. In the above equations, the mass scaling factor $d$ is given by
\begin{eqnarray}
d^2=\frac{m_X}{\mu}\frac{2m_{\mathrm{H}}}{2m_{\mathrm{H}}+m_X}.
\end{eqnarray}
The three-body reduced mass is defined as follows to preserve the phase-space volume element~\cite{Delves}:
\begin{eqnarray}
\mu=\sqrt{\frac{m_{\mathrm{H}}^2 m_X}{2m_{\mathrm{H}}+m_X}}.
\end{eqnarray}

In the adiabatic hyperspherical representation, the hyperradius $R$,
$R^2$=$\rho_{12}^2+\rho_{12,3}^2$,
is the only coordinate with the dimension of length and represents the overall size of the three-body system.
The remaining degrees of freedom, the hyperangles, are represented collectively by $\Omega$. We use body-frame Delves' coordinates~\cite{LinReview} 
such that $\Omega\equiv(\phi,\theta,\alpha,\beta,\gamma)$, with
\begin{eqnarray}
\phi=\tan^{-1}\left(\frac{\rho_{12,3}}{\rho_{12}}\right), \qquad 0\leq\phi\leq\pi/2;
\end{eqnarray}
and $\theta$, the angle between the vectors $\boldsymbol{\rho}_{12}$ and $\boldsymbol{\rho}_{1,23}$ such that $0\leq\theta\leq\pi$. The remaining hyperangles are the three Euler angles $\alpha$, $\beta$ and $\gamma$ describing
the rotation of the plane containing the three particles. 
As a result, the interparticle distances $r_{ij}$ are determined in terms of the internal coordinates ($R$, $\theta$, $\phi$) only:
\begin{eqnarray}
\!\!\!r_{12}&\!\!\!=& \!\!\! R d\cos\phi\\
\!\!\!r_{23}&\!\!\!=& \!\!\!R(\frac{d^2}{4}\cos^2\phi\!\!+\!\!\frac{1}{d^2}\sin^2\phi\!+\!\frac{1}{2}\sin2\phi\cos\theta)^{1/2},\\
\!\!\!r_{31}&\!\!\!=& \!\!\!R(\frac{d^2}{4}\cos^2\phi\!\!+\!\!\frac{1}{d^2}\sin^2\phi\!-\!\frac{1}{2}\sin2\phi\cos\theta)^{1/2}.
\end{eqnarray}
This definition of the hyperangles facilitates the symmetrization of the wave function under exchange of the two H atoms.

After rescaling the three-body wave function $\Psi$ as $\psi$=$R^{5/2}\Psi$, the three-body Schr{\"o}dinger equation takes the form
\begin{eqnarray}
\left[-\frac{1}{2\mu}\frac{\partial^2}{\partial R^2}+\frac{\Lambda^2}{2\mu R^2}+V(R,\Omega) \right]\psi=E\psi,
\label{Eq_Sch}
\end{eqnarray}
where $V(R,\Omega)$ includes all the interactions and $\Lambda^2$ is the hyperangular momentum operator, defined by taking $\boldsymbol{\rho}_{12}$ to be the quantization axis for the body-fixed frame~\cite{LinReview} and expressed as:
\begin{eqnarray}
\Lambda^2=T_0+T_1+T_2-1/4,
\end{eqnarray}
with
\begin{eqnarray}
T_0&=&\frac{\partial^2}{\partial \phi^2}-\frac{4}{\sin^2 2\phi}\frac{1}{\sin\theta}\frac{\partial}{\partial\theta}\left(\sin\theta\frac{\partial}{\partial\theta} \right),\\
T_1&=&\frac{4}{\sin^2 2\phi}\frac{1}{\sin^2\theta}J_z^2-\frac{1}{\cos^2\phi}(2J_z^2-J^2),\\
T_2&=&\frac{1}{\cos^2\phi}\left(2i J_y\frac{\partial}{\partial\theta}+2\cot\theta J_xJ_z\right).
\end{eqnarray}
The components {$\boldsymbol{J}\equiv(J_x,J_y,J_z)$} are the total orbital angular momentum operator projected on the body-frame axes.

\begin{figure}
\includegraphics[clip=true,scale=0.65]{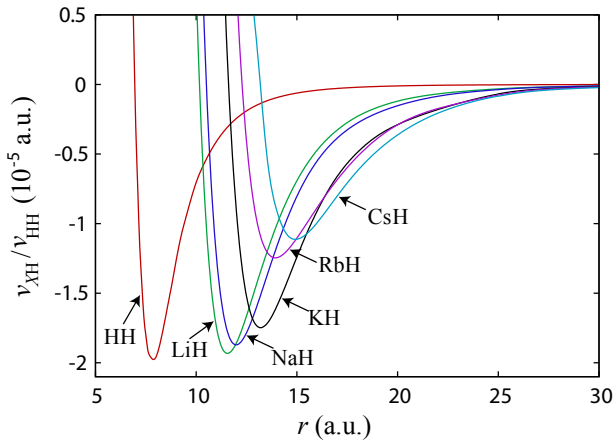}
\caption{The $^3\Sigma_u$ potentials for H+H and $X$+H. Among all these combinations, only $X$+H systems have a single, weakly bound molecular state.}
\label{Fig_2BPot}
\end{figure}

Since we assume the atoms to be spin-stretched, i.e., in the total spin state with the largest magnitude spin projection, the relevant Born-Oppenheimer potential surface is the lowest
quartet surface. We approximate this surface as a pairwise sum of $^3\Sigma_u$ two-body potentials:
\begin{eqnarray}
V(R,\Omega)=v_{\rm HH}(r_{12})+v_{X{\rm H}}(r_{23})+v_{X{\rm H}}(r_{31}).
\label{PairSum}
\end{eqnarray}
The two-body potentials $v_{\rm HH}(r)$ and $v_{X{\rm H}}(r)$ are shown in Fig.~\ref{Fig_2BPot}. 
At small distances, these potentials are determined from {\it ab initio} calculations~\cite{Kolos,Cote,Cote1} 
while their long-range behavior is determined by the usual dispersion potentials~\cite{Bromley,Cote}.

Note that for three atoms, a non-additive three-body term should be included in $V(R,\Omega)$. 
This three-body term depends on the spatial configuration of the three atoms and can be significant for certain configurations. 
The only fully quantum mechanical recombination calculation for a realistic system that has so far included the three-body term found its effect to be negligible~\cite{SunoHe2}. 
But, that work treated He, and He is not very polarizable compared to H or the other alkalis.  The impact of the
three-body term for the present systems is thus expected to be correspondingly larger.
Unfortunately, neither the full three-body surface nor the three-body term is available for the quartet state of H+H+$X$. 
The importance of the three-body term for the systems we are investigating, though, can be estimated qualitatively by looking at the available three-body terms  
for the quartet surface of identical alkali atoms~\cite{ThreebodyTerm1,ThreebodyTerm2}, as they have similar electronic structure. 
The three-body term is most significant when all the atoms are close together. 
For Li, the three-body potential can make the minimum of the total potential about four times deeper than the pair-wise sum potential~\cite{ThreebodyTerm2}. 
For heavier alkali atoms, the three-body term can change the potential minimum by a factor of 1.2--1.5~\cite{ThreebodyTerm1}. 
When the atoms are far apart, a three-body dispersion interaction should also be included. The contribution of this interaction, however, is much smaller. 
For Li, the three-body dispersion interaction is a few percent of the pair-wise sum potential, and even smaller for heavier alkali atoms~\cite{ThreebodyTerm2}. 
We thus expect that our results will change dramatically when the three-body interactions are included. 
We know, for instance, that the numerical value of the rate can change over a broad range for model problems~\cite{JosePhase}.
However, as the first calculation for H+H+$X$ systems, our results can give a sense of the order of magnitude for the three-body observables, 
and serve as a starting point for the study of three-body interactions in such three-body systems. 
We further note that including a three-body term or a full three-body surface poses no particular problem for our approach~\cite{JosePhase}.

In order to solve Eq.~(\ref{Eq_Sch}), we first expand the three-body wave function as
\begin{eqnarray}
\psi=\sum_{\nu=0}^{\infty} F_{\nu E}(R)\Phi_\nu(R;\Omega),
\end{eqnarray}
where $\Phi_\nu(R;\Omega)$ are the channel functions obtained as the solutions of the adiabatic equation
\begin{eqnarray}
\left[\frac{\Lambda^2}{2\mu R^2}+V(R,\Omega)\right]\Phi_\nu=U_\nu(R)\Phi_\nu.
\label{Eq_Adiabatic}
\end{eqnarray}
We solve this equation as a function of $R$,  and its eigenvalues are the adiabatic potentials $U_{\nu}(R)$.
Therefore, upon substitution of $\psi$, Eq.~(\ref{Eq_Sch}) reduces to a set of coupled  ordinary differential equations:
\begin{eqnarray}
&&\!\!\!\!\!\!\!\!\left[-\frac{1}{2\mu}\frac{d^2}{d R^2}+U_{\nu}(R)\right]\!F_{\nu E}(R)
\!-\!\frac{1}{2\mu}\sum_{\nu'}\!\left[P_{\nu \nu'}(R)\frac{d}{dR}\right.\nonumber\\
&&\!\!\!+\frac{d}{dR}P_{\nu \nu'}(R)+Q_{\nu \nu'}(R)\bigg]F_{\nu' E}(R)\!=\!E F_{\nu E}(R),
\label{Eq_Radial}
\end{eqnarray}
with non-adiabatic couplings $P_{\nu \nu'}$ and $Q_{\nu \nu'}$ given by
\begin{eqnarray}
P_{\nu \nu'}(R)=\left\langle\!\!\!\left\langle\Phi_{\nu}\left|\frac{d}{d R}\right|\Phi_{\nu'}\!\right\rangle\!\!\!\right\rangle,\\
Q_{\nu \nu'}(R)=\left\langle\!\!\!\left\langle\frac{d\Phi_{\nu}}{d R}\left|\frac{d\Phi_{\nu'}}{d R}\right.\!\!\right\rangle\!\!\!\right\rangle.\label{Couplings}
\end{eqnarray}
Here, the double brackets denote integration over only the hyperangular degrees of freedom.

In our calculations, the biggest computational burden comes from solving the five-dimensional adiabatic equation~(\ref{Eq_Adiabatic}). To facilitate its solution, we
separate out the external degrees of freedom ($\alpha$, $\beta$, $\gamma$) and simultaneously obtain eigenstates of total orbital angular
momentum and parity by further expanding the adiabatic wave functions $\Phi_\nu$ on the basis of symmetrized Wigner $D$ functions~\cite{LinReview}:
\begin{eqnarray}
\Phi_\nu(R;\Omega)=\sum_{K=0}^J u_{\nu K}(R;\theta,\phi)\tilde{D}_{KM}^{J\Pi}(\alpha,\beta,\gamma),\label{ChannelFunctions}
\end{eqnarray}
where $\Pi$ denotes the total parity of the system, and $K$ and $M$ denote the projection of the total orbital angular momentum on the body-fixed and space-fixed $z$ axes, respectively.
Since molecules $X$H have only a single $s$-wave state, recombination only happens for the parity-favored case, i.e., when $\Pi$=$(-1)^J$.

Although we are solving for the motion of the nuclei, we must require that the total wave function, including the electronic degrees of freedom, is antisymmetric under exchange of the two protons.
For the electronic symmetry we are considering, exchanging protons introduces a sign change in the electronic wave function. Thus, the nuclear part of the wave function must be symmetric under 
proton exchange. 

Since we neglect hyperfine interactions,  we can couple the
two protons' spin to give a total spin $I$=0,1 and consider  
their contribution to recombination independently.
For $I$=1, the spin wave function is symmetric under exchange, requiring the spatial wave function to also be symmetric. Such spatial symmetry
leads to non-vanishing $K_3$ at ultracold temperatures~\cite{WignerLaw}. For $I$=0, however, the spin wave function is antisymmetric under exchange of the two protons, and
the spatial wave function is therefore also antisymmetric. For this symmetry, $K_3$ vanishes in the zero temperature limit~\cite{WignerLaw}. 
In the present work, we calculate $K_3$ for $I$=1, which 
is the dominant recombination process for ultracold temperatures.  For more general cases where the $I$=1 state is not preferentially prepared, our results give only a partial contribution to $K_3$ for temperatures beyond the ultracold regime.

We built the exchange symmetry of the two protons into the boundary conditions of the body-frame components $u_{\nu K}$.
Permuting the two protons only affects the hyperangles $\Omega$:
\begin{eqnarray}
P_{12}\tilde{D}_{KM}^{J\Pi}&=& \Pi(-1)^K \tilde{D}_{KM}^{J\Pi},\\
P_{12}\theta&=&\pi-\theta.
\end{eqnarray}
For even parity, the permutation requirements can be equivalently expressed as $u_{\nu K}$ being symmetric about $\theta$=$\pi/2$ for even $K$ and antisymmetric for odd $K$. 
For odd parity, $u_{\nu K}$ should be antisymmetric for even $K$ and symmetric for odd $K$.  Imposing these boundary conditions, we need only solve Eq.~(\ref{Eq_Adiabatic}) in the range $0\leq\theta\leq\pi/2$.

Asymptotically, i.e., as $R\rightarrow \infty$, the adiabatic potentials with the diagonal couplings included are determined by the energies of the break-up components.
For the atom-molecule channel, the potentials behave like
\begin{eqnarray}
W_0(R)=U_0(R)-\frac{1}{2\mu}Q_{0,0}\rightarrow E_{X\mathrm{H}}+\frac{l(l+1)}{2\mu R^2},
\label{atomdimerchannel}
\end{eqnarray}
where the partial angular momentum $l$ is the relative orbital angular momentum between the atom and the molecule. 
Since all of the $X$H systems have only an $s$-wave bound state,
$l$=$J$.
For the three-body break-up channels, the potentials behave like 
\begin{eqnarray}
W_\nu(R)\rightarrow \frac{\lambda(\lambda+4)+15/4}{2\mu R^2}.
\label{threebodychannels}
\end{eqnarray}
The values of $\lambda$ are non-negative integers determined by $J^\Pi$ and the identical particle symmetry~\cite{WignerLaw}. 

Accurate numerical calculations of the three-body observables depend largely on the accuracy of the adiabatic potentials and channel functions [Eqs.~(\ref{Eq_Adiabatic}) and (\ref{ChannelFunctions})] and, 
ultimately, on the non-adiabatic couplings [Eq.~(\ref{Couplings})].
By expanding the body-frame components $u_{\nu K}(R;\theta,\phi)$ on a two-dimensional, direct product B-spline basis~\cite{Bspline}, we obtain accurate potentials and couplings up to
$R\approx 2000$~a.u.. Beyond this distance, we extrapolate the potentials using the known asymptotic expansions~\cite{NielsenRev}.
Typically, a ($\theta$, $\phi$) mesh of 60$\times$250 gives eigenvalues converged to at least eight digits. We have found that due to the sharp avoided crossings occurring
at small $R$, a hyperradial grid of about 3000 points is necessary to accurately resolve most of the abrupt changes in the non-adiabatic couplings. Many sharper crossings remain, though, that must be traced individually.

\section{Three-body scattering observables}

It is well known~\cite{BraatenReview} that when the scattering length $a$ greatly exceeds the characteristic range of the two-body interaction, 
three-body scattering observables are dramatically affected.
For the systems we consider here, the long-range part of the two-body interaction is the van der Waals potential $-C_6/r_{ij}^6$. Therefore, 
they are characterized by the van der Waals length ${l}_{vdW}=(2\mu_{ij}C_6)^{1/4}$~\cite{BraatenReview}, where $\mu_{ij}$ is the two-body reduced mass.
In Table~\ref{Tab_2BScatt}, we list the bound state energies, the scattering lengths, and the van der Waals lengths
for all of the two-body potentials we used. 
Notice that none the scattering lengths are substantially larger than the van der Waals lengths and thus the condition for universal behavior ($|a|\gg l_{vdW}$) is not fulfilled. 
As a result, we do not expect to observe universal physics for these systems.

\begin{table}
\begin{ruledtabular}
\begin{tabular}{cccc}
& $E_{X{\rm H}}$ (a.u.) & $a$ (a.u.) & ${l}_{vdW}$ (a.u.)\\
\hline
H+H & --- & 1.557 & 10.45\\
Li+H&-1.268$\times$$10^{-7}$ & 63.71 & 21.50\\
Na+H&-3.376$\times$$10^{-7}$ & 43.26 & 22.58\\
K+H &-7.360$\times$$10^{-7}$ & 34.72 & 25.12\\
Rb+H&-2.446$\times$$10^{-7}$ & 50.24 & 25.92\\
Cs+H&-1.784$\times$$10^{-7}$ & 56.85 & 27.18
\end{tabular}
\end{ruledtabular}
\caption{The two-body bound state energy $E_{X{\rm H}}$, scattering length $a$, and van der Waals length $l_{vdW}$ for the H+H and $X$+H interactions.}
\label{Tab_2BScatt}
\end{table}

After obtaining the potentials and couplings, we solve the hyperradial equation (\ref{Eq_Radial}) using finite elements as described in Ref.~\cite{BurkeThesis}. 
For recombination processes, the total recombination rate $K_3$ is the sum over all the partial wave contributions $K_3^{J\Pi}$~\cite{SunoHe1,SunoHe2}:
\begin{eqnarray}
K_3=\sum_{J,\Pi}K_3^{J\Pi}=2!\sum_{J,\Pi}\sum_i\frac{32(2J+1)\pi^2}{\mu k^4}\left|S_{f\leftarrow i}^{J\Pi} \right|^2,
\end{eqnarray}
where $k$=$\sqrt{2\mu E}$ and $S_{f\leftarrow i}^{J\Pi}$ is the scattering matrix element from the initial three-body
continuum channel to the final atom-molecule channel. From the asymptotic form of the three-body entrance channel, 
the threshold behavior of $K_3^{J\Pi}$ is determined by the smallest $\lambda$ for that symmetry~\cite{WignerLaw} such that:
\begin{eqnarray}
K_3^{J\Pi}\propto E^{\lambda_{\mathrm{min}}^{J\Pi}}.
\end{eqnarray}
In our calculations, we have included the lowest six partial-waves $J^\Pi$, implying that $\lambda_{\rm min}^{J\Pi}$=0, 1, 2, 3, 4 and 5, respectively. 

Collision-induced dissociation H+$X$H$\rightarrow$$X$+H+H is the time reversed process of three-body recombination. The dissociation rate $D_3$ is defined as~\cite{SunoHe1}
\begin{eqnarray}
D_3=\sum_{J,\Pi}D_3^{J\Pi}=\sum_{J,\Pi}\sum_f\frac{(2J+1)\pi^2}{\mu_{12,3} k_{12,3}}\left|S_{f\leftarrow i}^{J\Pi} \right|^2,
\end{eqnarray}
where $k_{12,3}=\sqrt{2\mu_{12,3} (E-E_{X{\rm H }})}$ and $\mu_{12,3}=m_{\rm H} (m_{\rm H}+m_{X})/(2m_{\rm H}+m_{X})$ is the reduced mass between the H atom and the $X$H molecule.

Note that the channels that the indices $i$ and $f$ refer to are reversed from those in $K_3$.  
Since the $S$-matrix is unitary, $D_3$ can be readily calculated once 
the $S$-matrix elements for $K_3$ are known.
Near the three-body breakup threshold where collision-induced dissociation becomes
energetically possible, $D_3^{J\Pi}$ behaves like~\cite{WignerLaw}:
\begin{eqnarray}
D_3^{J\Pi}\propto E^{\lambda_{\mathrm{min}}^{J\Pi}+2}.
\label{Eq_D3Thresh}
\end{eqnarray}

For atom-molecule collisions, the elastic cross section is~\cite{SunoHe2}
\begin{eqnarray}
\sigma_2=\sum_{J,\Pi}\sigma_2^{J\Pi}=\sum_{J,\Pi}\frac{(2J+1)\pi}{k_{12,3}^2}\left|S_{0\leftarrow 0}^{J\Pi}-1\right|^2,
\end{eqnarray}
The threshold behavior of $\sigma_2^{J\Pi}$, in contrast to recombination, is determined solely by $J$ and
follows the standard Wigner threshold law, 
\begin{eqnarray}
\sigma_2^{J\Pi}\sim (E-E_{X{\rm H}})^{2J}.
\end{eqnarray}

The calculation of scattering solutions to Eq.~(\ref{Eq_Radial}) are complicated by the sharp avoided crossings in the adiabatic potentials: we typically use 5$\times$10$^4$ hyperradial elements 
distributed as $R_i\propto i^3$ from $R$=10 a.u. to $R\approx 2000$ a.u.. 
In the asymptotic region ($R>2000$~a.u.), the density of elements is fixed to eight elements per shortest de Broglie wavelength.
To calculate the scattering observables, we match 
the numerical solutions to the asymptotic analytical solutions at $R$=$10^5$~a.u. for recombination, and at $R$=5$\times$$10^3$~a.u.
for atom-molecule collisions. 
The convergence of the scattering observables with respect to the number of adiabatic channels is also dramatically affected by the sharp avoided crossings. 
Even the threshold behavior for $K_3^{J\Pi}$ and $\sigma_2^{J\Pi}$ requires
a fairly large number of adiabatic channels for convergence, which we take to be from 12 to 25 for all the calculations. 
The resulting $K_3$ and $\sigma_2$ are converged to at least two digits for all partial waves, and the three-body bound state energies are converged to three digits.

In our calculations, we have included $J^\Pi$=$0^+$, $1^-$, $2^+$, $3^-$, $4^+$ and $5^-$ for the convergence of the total rates and cross sections at high energies. 
The overall convergence of the total total rates and cross sections are converged to two digits for $E<200$~mK and one digit for 
200~mK$<E<$500~mK.
\section{results}
\label{results}

\subsection{Three-body recombination rates}

\begin{figure}
\includegraphics[clip=true,scale=0.65]{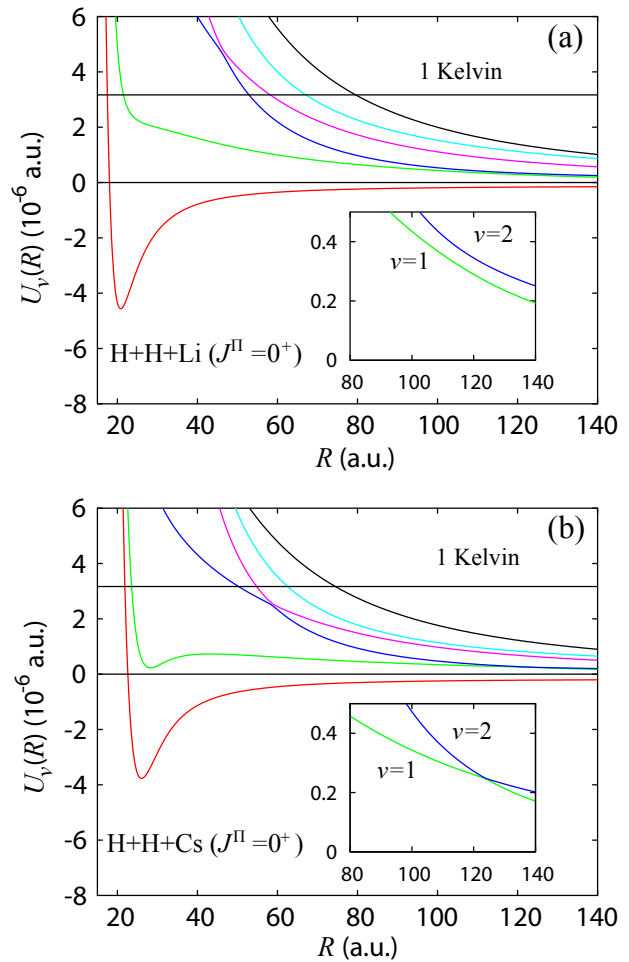}
\caption{(color online) The adiabatic three-body potentials for (a) H+H+Li, and (b) H+H+Cs. 
The insets show the avoided crossings within the energy range we have considered for the rates and cross sections.}
\label{Fig_3BPot}
\end{figure}

Since the adiabatic hyperspherical potentials $U_{\nu}(R)$ are important in understanding the underlying three-body physics involved in the scattering processes, 
we first discuss their behavior.
We see that the avoided crossings become sharper as we go to heavier alkali atoms. 
As an illustration, in Fig.~\ref{Fig_3BPot} we show the lowest six adiabatic potentials $U_\nu(R)$ with $J^\Pi$=$0^+$ for 
H+H+Li and H+H+Cs. The potentials for higher partial waves behave similarly but become more repulsive as $J$ increases.

\begin{figure}
\includegraphics[clip=true,scale=0.65]{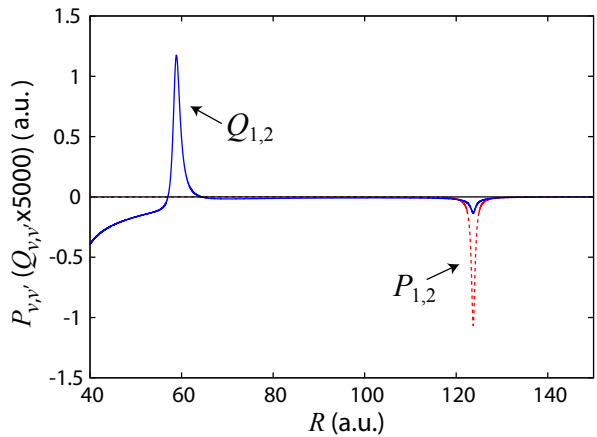}
\caption{(color online) The non-adiabatic couplings  for H+H+Cs with $J^\Pi$=$0^+$. Note that $Q_{1,2}$ has been multiplied
by 5000 to have a magnitude similar to $P_{1,2}$.
The index $\nu$ for the adiabatic potentials starts from 0.}
\label{Fig_Couplings}
\end{figure}

To demonstrate the effects of sharp avoided crossings on the adiabatic potentials on the non-adiabatic couplings,
we show a key crossing in the insets of Fig.~\ref{Fig_3BPot}.  Figure~\ref{Fig_Couplings} shows the corresponding
couplings  $P_{\nu\nu'}$ and $Q_{\nu\nu'}$ 
for H+H+Cs with $J^\Pi$=$0^+$. It can be seen that when a sharp avoided crossing occurs between two 
potential curves, the  $P_{\nu \nu'}$ and $Q_{\nu \nu'}$ coupling those curves show sharp spikes at that $R$.  These
couplings must be carefully traced out with a dense hyperradial grid in order to obtain an accurate solution of Eq.~(\ref{Eq_Radial}).

In our calculations, the convergence of $K_3$ depends critically on the behavior of the adiabatic potentials. 
We have found that, except for H+H+Li, all the systems have sharp avoided crossings below 1 Kelvin. 
To calculate $K_3$ for energies above the crossings like the one shown in the inset of Fig.~\ref{Fig_3BPot}(b), 
both of the potentials involved in the crossing need to be included to avoid spurious resonances.

\begin{figure*}
\includegraphics[clip=true,scale=0.65]{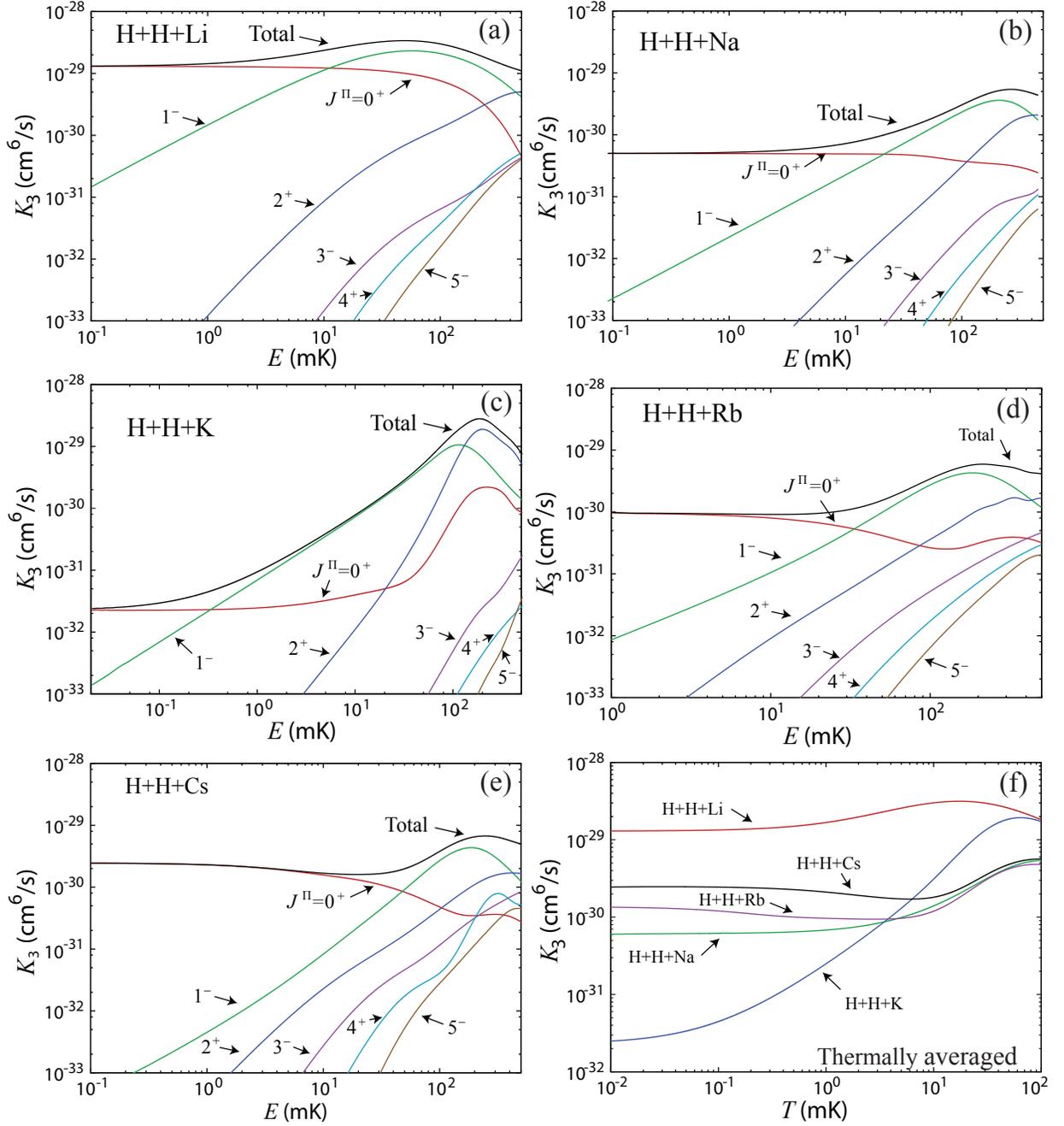}
\caption{(color online) The total three-body recombination rate $K_3$ and the partial rates $K_3^{J\Pi}$ for H+H+$X$$\rightarrow$H+$X$H, 
where $X$ is (a) Li, (b) Na, (c) K, (d) Rb and (e) Cs. The thermally-averaged total recombination rates are shown in (f).
}
\label{Fig_Rates}
\end{figure*}

The three-body recombination rates for different alkali species $X$ are shown in Fig.~\ref{Fig_Rates}(a)--(f). The corresponding data are available
in electronic form~\cite{Data}.
Generally, the three-body recombination rates are dominated by the $J^{\Pi}=0^+$ contribution at ultracold energies, 
and by the $J^{\Pi}=1^-$ and $2^+$ contributions near the highest energies we have calculated.
It can be seen that the total three-body recombination rates for H+H+Na,  H+H+Rb, and H+H+Cs behave similarly and that 
the rates for H+H+Li and H+H+K recombination behave differently. 
In particular, the $J^\Pi=0^+$ partial rates for H+H+K recombination are much smaller than the rates for other systems near the zero-energy threshold. 
The total rates for H+H+K recombination are then dominated by the $J^\Pi=1^-$ partial wave contribution for a large energy range from about 0.5 mK to 50 mK.
Interestingly, we have observed that the threshold values of the recombination rates for different $X$ are ordered by the magnitude of their non-adiabatic 
couplings $P_{01}$ and $Q_{01}$ at large hyperradii $R>200$~a.u. for $0^+$. This suggests that for the present cases $0^+$ recombination is dominated 
by inelastic transitions from the lowest continuum channel to the atom-molecule channel at large distances.

In thermal gases, it is crucial to consider the thermal distribution of the collisional energies
when calculating a collision rate. Assuming a Boltzmann distribution, we have performed a
thermal average of the energy-dependent rates. The thermally averaged recombination rate $\left<K_3\right>$ are given by~\cite{ThermAvg}:
\begin{equation}
\left<K_3\right>=\frac{1}{2(k_B T)^{3}}\int_0^\infty K_3 E^{2} e^{-E/k_B T} dE.
\label{Eq_K3ThermAvg}
\end{equation}
The results are shown in Fig.~\ref{Fig_Rates}(f). 
To perform the thermal average, we extrapolate $K_3$ from the lowest energy we have calculated to zero energy using the known threshold behavior.
Since we could not similarly extrapolate to infinite energy for the integral in Eq.~(\ref{Eq_K3ThermAvg}),
the integral was limited to the energies we could calculate.  Consequently,
 the thermally averaged rates are converged to more than one digit only for temperatures below 100 mK.
It can be seen that the energy dependence of the rates is largely preserved.
Further, the thermally averaged rates for these systems lie close to each other when the temperature is beyond 10 mK.
In Table~\ref{Tab_Rates}, we list the values of $K_3$ for the processes H+H+$X$$\rightarrow$$X$H+H in the zero-energy limit for reference.
\begin{table}
\begin{ruledtabular}
\begin{tabular}{llll}
& & $K_3$ (cm$^6$/s)&\\
\hline
&H+H+Li&1.3$\times$$10^{-29}$&\\
&H+H+Na&6.1$\times$$10^{-31}$&\\
&H+H+K&2.3$\times$$10^{-32}$&\\
&H+H+Rb&9.7$\times$$10^{-31}$&\\
&H+H+Cs&2.5$\times$$10^{-30}$&\\
\end{tabular}
\end{ruledtabular}
\caption{The zero-energy limit of the three-body recombination rates $K_3$.}
\label{Tab_Rates}
\end{table}

\subsection{Collision induced dissociation rates}
\begin{figure}
\includegraphics[clip=true,scale=0.65]{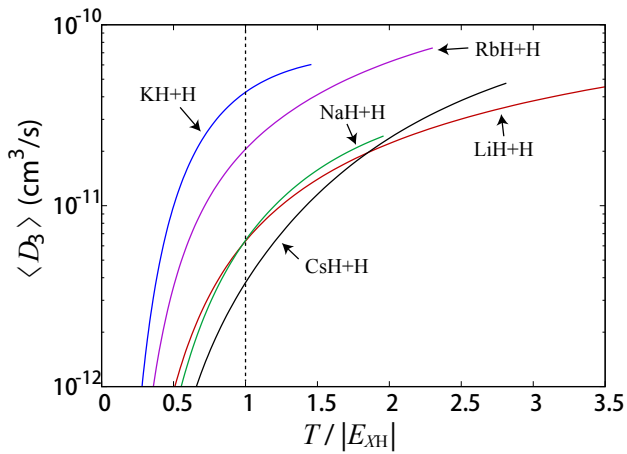}
\caption{(color online) The thermally averaged total collision induced dissociation rate $\left<D_3\right>$ for H+$X$H$\rightarrow$$X$+H+H. For all species, the rates are shown up to 100~mK beyond the three-body break-up threshold which is indicated by the vertical dashed line.
}
\label{Fig_D3_Therm}
\end{figure}

Using the simple relation between $K_3$ and $D_3$, we have also calculated $D_3$ for the same range of $E$. 
In Fig.~\ref{Fig_D3_Therm}, we show the thermally averaged collision-induced dissocation rate $\left<D_3\right>$ as a function of 
the temperature of the $X$H+H mixture, 
where $\left<D_3\right>$ is given by~\cite{BurkeThesis}
\begin{equation}
\left<D_3\right>=\frac{2}{\sqrt{\pi}}\frac{1}{(k_B T)^{3/2}}\int_0^\infty D_3 E^{1/2} e^{-E/k_B T} dE.
\end{equation}
The energy $E$ in the integrand is relative to the $X$H+H threshold.  That is, it is the $X$H+H scattering
energy, and this must be taken into account when evaluating $D_3(E)$.
It is interesting to note that although dissociation is allowed only when the collision energy exceeds the molecular binding energy, Fig.~\ref{Fig_D3_Therm} shows that in a thermal gas dissociation can occur for temperatures well below the dissociation threshold.
In fact, because we know the threshold behavior from Eq.~(\ref{Eq_D3Thresh}) to be $D_3\propto(E-E_{X{\rm H}})^2$ for
$E\ge E_{X{\rm H}}$ (and zero below $E_{X{\rm H}}$), we can explicitly calculate $\left<D_3\right>$ below threshold:
\begin{equation*}
\left<D_3\right>\propto 2\sqrt{x} (15x-2) e^{-1/x}+\sqrt{\pi}[4+3x(5x-4){\rm erfc}(1/\sqrt{x})]
\end{equation*}
with $x=k_BT/E_{X{\rm H}}$.  By contrast, $\left<K_3\right>$ and $\left<\sigma_2\right>$ 
have the same threshold behavior as the energy-dependent quantities (with $E$ replaced by $k_BT$).  This formula for
$\left<D_3\right>$ is likely valid only for temperatures below $E_{X{\rm H}}$ since the tail of the thermal distribution starts
sampling energies outside the threshold regime for higher temperatures, making our assumption for the behavior of $D_3$ invalid.

\subsection{Atom-molecule elastic cross sections}

\begin{figure}
\includegraphics[clip=true,scale=0.65]{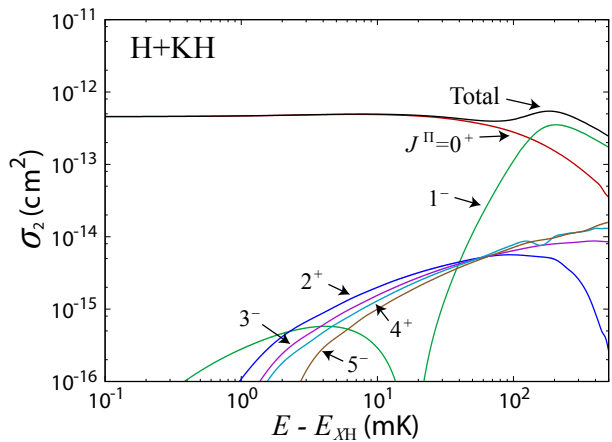}
\caption{(color online) The total atom-molecule elastic cross section $\sigma_2$ and the partial cross sections $\sigma_2^{J\Pi}$ for H+KH$\rightarrow$H+KH.}
\label{Fig_Cross_K}
\end{figure}

\begin{figure}
\includegraphics[clip=true,scale=0.65]{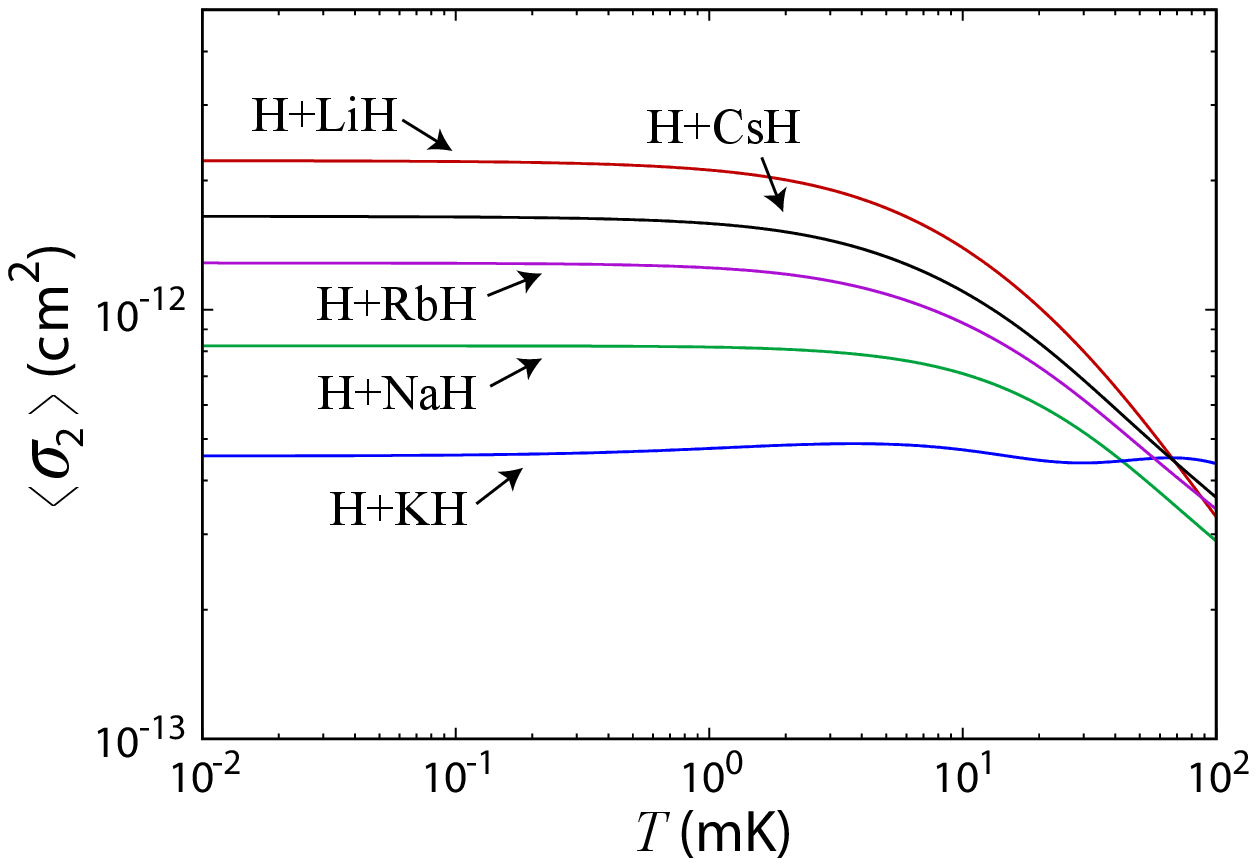}
\caption{(color online) The thermally averaged total atom-molecule elastic cross section $\left<\sigma_2\right>$ for H+$X$H$\rightarrow$H+$X$H.}
\label{Fig_Cross_Therm}
\end{figure}

As a representative example, we plot the total and partial cross sections for elastic collisions between H and KH in Fig.~\ref{Fig_Cross_K}. 
The $J^\Pi$=$0^+$ partial wave contribution dominates for collisional energies
below 100 mK, beyond which the $J^\Pi$=$1^-$ contribution becomes dominant. The $J^\Pi$=$1^-$ contribution has a pronounced minimum near 20 mK, but this feature
has only a negligible effect on the total cross section. The partial atom-molecule elastic cross sections for the other alkali species are not shown, as their energy-dependence
is qualitatively the same as shown for H+KH. Instead, we show in Fig.~\ref{Fig_Cross_Therm} the thermally averaged cross sections for all the alkali species.
The thermally averaged cross sections can be derived from the  thermally averaged elastic scattering rate, and is given by~\cite{BurkeThesis} 
\begin{equation}
\left<\sigma_2\right>=\frac{1}{(k_B T)^2}\int_0^\infty \sigma_2 E e^{-E/k_B T} dE.
\end{equation}
The total elastic cross sections for all alkali species are converged to two digits for all energies.
All these data are available in electronic form~\cite{Data}.

In Table~\ref{Tab_ScattL}, we list the values of the elastic cross section extrapolated to zero temperature as well as the 
corresponding values for the atom-molecule scattering length $a_{\mathrm{H}+X\mathrm{H}}$ for all the alkali species. 
Both quantities increase with the respective values of the two-body scattering lengths (see Table~\ref{Tab_2BScatt}), 
or equivalently, the size of the molecular state. In fact, we can get an order-of-magnitude estimate for the elastic cross-section by simply using 
$\sigma_2\approx4\pi a^2$. These estimated values are also shown in Table~\ref{Tab_ScattL}. 
The rough agreement implies that for our simple pair-wise sum potential the zero-energy
elastic cross section is mainly determined by the size of the $X$H molecule.  By extension,
the atom-molecule scattering length can be approximated by the $X$+H scattering length at the
same level of approximation.

\begin{table}
\begin{ruledtabular}
\begin{tabular}{lllccl}
& &$\sigma_2$ (cm$^2$)&$4\pi a^2$ (cm$^2$)&$a_{\mathrm{H}+X\mathrm{H}}$ (a.u.)&\\
\hline
&H+LiH&2.2$\times 10^{-12}$&1.4$\times 10^{-12}$&80&\\ 
&H+NaH&8.2$\times 10^{-13}$&6.6$\times 10^{-13}$&48&\\
&H+KH&4.6$\times 10^{-13}$&4.2$\times 10^{-13}$&36&\\
&H+RbH&1.3$\times 10^{-12}$&8.9$\times 10^{-13}$&60&\\
&H+CsH&1.6$\times 10^{-12}$&1.1$\times 10^{-12}$&68&\\
\end{tabular}
\end{ruledtabular}
\caption{The atom-molecule zero temperature elastic cross-section and scattering length between H and $X$H.}
\label{Tab_ScattL}
\end{table}

\subsection{Three-body bound state energies}

To complete our study of these systems, we calculate the three-body bound states. 
The three-body energy spectra for the H+H+Alkali systems are very simple due to their weakly-interacting nature. In our calculations, we have found only one
triatomic vibrational bound state for $J^\Pi=0^+$ for all the systems. No bound levels are found for higher angular momenta.
The three-body binding energies, relative to the atom-molecule break-up threshold, are listed 
in Table~\ref{Tab_E3b}. 

To get a sense of the sizes of the triatomic molecules, we have also calculated the expectation values of the interatomic distances 
$\left<r_{X{\rm H}}\right>$ and $\left<r_{\rm HH}\right>$, given by
\begin{equation}
\left<r_{A{\rm H}}\right>=\sum_{\nu,\nu'}\int_{0}^\infty F_\nu(R)F_{\nu'}(R)\left<\!\left<\Phi_\nu|r_{A{\rm H}}|\Phi_{\nu'}\right>\!\right>d R,
\end{equation}
where $A$ represents $X$ or H.  From these bond lengths, we can also calculate the bond angle at the $X$ atom and
find them to be consistently around 100$^\circ$ for all species.
All of this geometrical information is included in Table~\ref{Tab_E3b}.  From the small binding energies and
large bond lengths, it is clear that these are very floppy states as is expected for van der Waal's molecules.
We expect, though, that the inclusion of three-body terms in the interaction potential will tend to bind these
states more strongly, reducing the bond lengths correspondingly.  The three-body term may further tend to increase
the bond angle towards a linear configuration.  The three-body term might even be sufficient to bind additional
states, at least for some species.
\begin{table}
\begin{ruledtabular}
\begin{tabular}{lllccl}
& & $E_{X{\rm H}_2}$ (a.u.)&$\left<r_{X{\rm H}}\right>$ (a.u.)&$\left<r_{\rm HH}\right>$ (a.u.)& Bond angle\\
\hline
&LiH$_2$&9.02$\times$$10^{-8}$&43&65& 98$^\circ$\\
&NaH$_2$&2.58$\times$$10^{-7}$&30&46&100$^\circ$\\
&KH$_2$&6.24$\times$$10^{-7}$&25&37&95$^\circ$\\
&RbH$_2$&1.95$\times$$10^{-7}$&34&52&100$^\circ$\\
&CsH$_2$&1.43$\times$$10^{-7}$&38&59&102$^\circ$\\
\end{tabular}
\end{ruledtabular}
\caption{The $0^+$ triatomic bound state energies, expectation values of interatomic distances, and
bond angles.}
\label{Tab_E3b}
\end{table}

\begin{figure}[!h]
\includegraphics[clip=true,scale=0.65]{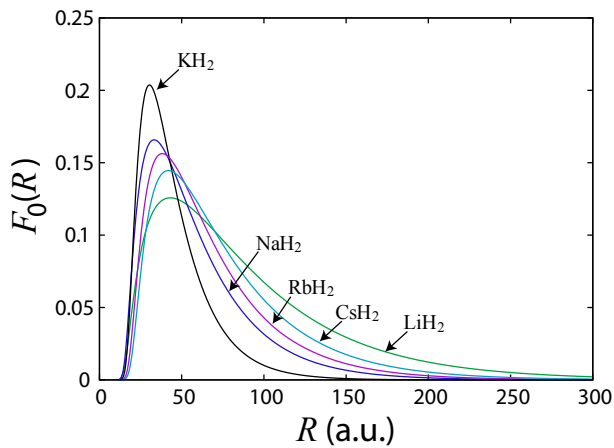}
\caption{(color online) The $J^\Pi=0^+$ hyperradial channel function $F_0(R)$ for the triatomic bound states.}
\label{Fig_FR}
\end{figure}

Finally, we have verified numerically that the triatomic binding energies are indeed predominantly determined by the lowest adiabatic channel $\nu=0$. 
Specifically, the channel probability $\int |F_\nu(R)|^2 dR$ for $\nu=0$ is
beyond 99\% for all the systems. The channel functions $F_0(R)$ are shown in Fig.~\ref{Fig_FR}. It is interesting to note that, 
except for KH$_2$, all the triatomic states have a large hyperradial extent, reaching values up to a few hundred atomic units 
consistent with the bond lengths listed in Table~\ref{Tab_E3b}.

\section{summary}
\label{summary}

In this paper, we have studied three-body scattering and the bound state spectra for two hydrogen atoms and one alkali atom using a fully quantum mechanical approach. 
Solving the three-body Schr{\"o}dinger equation in the adiabatic hyperspherical representation, we have calculated the three-body recombination rates and atom-molecule elastic cross sections
for all the alkali species for temperatures up to 0.5 Kelvin. The biggest uncertainty in our calculations by far is the interaction potential.
Nevertheless, we expect that our results give a correct order-of-magnitude estimate of the three-body scattering observables.
For three-body recombination, the lowest three partial waves dominate the total recombination rates in the energy range we have calculated. 
For the elastic atom-molecule collisions, the cross sections are dominated by a single partial wave
contribution for the energy range in our calculations, which is $J^\Pi$=$0^+$ at lower energies and $J^\Pi$=$1^-$ at higher energies. The bound state spectra are very simple,
with only one ro-vibrational three-body state for each of the alkali species.

Finally, the difficulty of sharply avoided crossings we met at small hyperradius raises an alert for doing adiabatic calculations for realistic systems. The complicated
short-range three-body dynamics can give rise to rapidly varying behavior in the adiabatic potentials and the non-adiabatic couplings, which makes the
adiabatic calculations much harder and less reliable. For such cases, a diabatic representation of some sort will become necessary, especially for small distances~\cite{DiabaticEsry,DiabaticWang}.

\begin{acknowledgments}
We acknowledge early assistance with the alkali-hydride potentials from J.J. Hua.  We are also 
grateful to R. C\^ot\'e and A. Derevianko for sharing their alkali-hydride potentials and data.
This work was supported in part by the National Science Foundation and in part by the Air Force Office of Scientific Research.
Y. W. and J.P.D. also acknowledge the support from the National Science Foundation under Grant No. PHY0970114.
\end{acknowledgments}


\begin{thebibliography}{99}
\bibitem{Data} The data for $K_3$ and $\sigma_2$ can be found in the APS data server.
\end{thebibliography}
\end{document}